\documentclass[11pt,a4paper]{article}
\pdfoutput=1
\usepackage{jheppub}
\newcommand{\beq}{\begin{equation}}
\newcommand{\eeq}{\end{equation}}

\usepackage{graphicx,subfigure,multirow,xcolor}
\def\bea{\begin{eqnarray}}
\def\eea{\end{eqnarray}}

\begin{document}

\title{Racetrack potentials and the de Sitter swampland conjectures}

\author[a,b]{Jose J. Blanco-Pillado}
\affiliation[a]{Department of Theoretical Physics, UPV/EHU, Bilbao, Spain}
\affiliation[b]{IKERBASQUE, Basque Foundation for Science, Bilbao, Spain}
\author[a]{Mikel A. Urkiola}
\author[a]{Jeremy M. Wachter}

\emailAdd{josejuan.blanco@ehu.eus}
\emailAdd{mikel.alvarezu@ehu.eus}
\emailAdd{jeremy.wachter@ehu.eus}

\def\changenote#1{\footnote{\bf #1}}

\abstract{We show that one can find de Sitter critical points (saddle points) in models of flux compactification of Type IIB String Theory without any uplifting terms and in the presence of several moduli. We demonstrate this by giving explicit examples following some of the ideas recently presented by Conlon in ref.~\cite{Conlon:2018eyr}, as well as more generic situations where one can violate the strong form of the de Sitter Swampland Conjecture. We stabilize the complex structure and the dilaton with fluxes, and we introduce a racetrack potential that fixes the K\"ahler moduli. The resultant potentials generically exhibit de Sitter critical points and satisfy several consistency requirements such as flux quantization, large internal volume, and weak coupling, as well as a form of the so-called Weak Gravity Conjecture. Furthermore, we compute the form of the potential around these de Sitter saddle points and comment on these results in connection to the refined and more recent version of the de Sitter Swampland Conjecture. }

\arxivnumber{1811.05463}
\maketitle

\section{Introduction}

It has recently been conjectured~\cite{Obied:2018sgi} that any potential consistent with quantum gravity should satisfy the bound
\beq\label{conjecture1}
    M_p \frac{|\nabla V|}{V} \ge c\,,
\eeq
where $c$ is a dimensionless constant of order unity. One of the obvious implications of this conjecture will be the impossibility of obtaining a metastable de Sitter vacua. However, the conjecture is much stronger than that: it also prohibits the existence of any critical point (saddle points) at positive values of the potential. This aspect of the conjecture is already under tension from several theoretical considerations that involve well known Standard Model physics~\cite{Denef:2018etk,Choi:2018rze,Murayama:2018lie}.  In particular, one would need to introduce some specific couplings of the Higgs field in order to satisfy eq.~(\ref{conjecture1}). 

String theory is a consistent theory of quantum gravity, and if the conjecture is correct, any potential obtained in this theory would have to satisfy eq.~(\ref{conjecture1}). This is particularly restrictive when we take into account the large number of 4-dimensional potentials that one could generate in the process of compactification from 10d. However, similar bounds on the potential have been found earlier in some string theory compactifications~\cite{Hertzberg:2007wc}. It is therefore very important to investigate whether or not this general statement about low energy effective field theory is true in a generic situation. This has been recently discussed in several papers, e.g., refs.~\cite{dS4.1,dS4.2,dS4.3,dS4.4,dS4.5,dS4.6,dS4.7,dS4.8,dS4.9,dS4.10,dS4.11,dS4.12,dS4.13,dS4.14,dS4.15}, mostly in relation to the validity of the approximations to obtain a 4-dimensional de Sitter space minima.

In this paper, we will focus our discussion on the existence of de Sitter saddle points in models of string compactification to demonstrate that the de Sitter conjecture is too restrictive. In fact, there is already some evidence in the literature for the existence of unstable de Sitter solutions, found from looking directly at the 10d equations of motion (see ref.~\cite{Andriot:2018ept} and references therein). This seems to suggest that one could find a 4d dimensional version of a potential that could bring the spacetime to this form, therefore violating the de Sitter conjecture given by eq.~(\ref{conjecture1}). However, it is not clear if those 10d solutions can be cast in a low energy effective field theory in 4d~\cite{Andriot:2018wzk}. Thus it is still necessary and interesting to look for some other possible counterexamples to the conjecture.

One can also take a different perspective and think about the purely 4d effective theory that one obtains from compactication. This approach has been taken in several of the most-studied examples of de Sitter vacua in string theory, in particular the well known KKLT model~\cite{Kachru:2003aw} or the Large Volume Scenario~\cite{LVS.1,LVS.2,LVS.3}. However, some of the ingredients in these constructions have been put into question by some authors~\cite{question.1,question.2,question.3,question.4} (see, however, ref.~\cite{Polchinski}). It is therefore interesting to ask whether one can find de Sitter critical points in these constructions that violate the de Sitter swampland conjecture with all of the ingredients well under control. This question has been addressed recently by Conlon in ref.~\cite{Conlon:2018eyr} in the context of moduli potentials.\footnote{See ref.~\cite{Olguin-Tejo:2018pfq} for another construction with de Sitter critical points.} His argument is based on the realization that in certain circumstances, the potential approaches zero from above along specific directions in field space, such as the internal volume or the dilaton. This fact, together with the existence of supersymmetric AdS minima in the interior of the moduli space, suggests that one should find a maximum of the potential somewhere between these two regions. This argument does not require the presence of any uplifting term in the effective potential and therefore seems more generic. In order to make the argument robust, one should show that the critical point is still there in the presence of several other fields, like the complex structure moduli and the other K\"ahler moduli. In this paper, we will show that this is indeed possible in constructions of Type IIB with several complex structure moduli, the dilaton, and one K\"ahler moduli. We further find that the presence of de Sitter saddle points is quite generic in these constructions, which include (but are not limited to) the cases discussed by ref.~\cite{Conlon:2018eyr}.

Furthermore, it has been argued in ref.~\cite{Moritz:2018sui} that a form of the Weak Gravity Conjecture~\cite{ArkaniHamed:2006dz} will inhibit the possibility of obtaining a viable model of compactication with a Racetrack Potential. In the following, we will show that the form of the parameters imposed by the authors in ref.~\cite{Moritz:2018sui} is in fact too restrictive, and one can find a set of coefficients that would not violate the Weak Gravity Conjecture.

The models we present here pass many requirements that one needs to impose to have some confidence on the results obtained from them. In particular we require that the following conditions are satisfied in our model:
\begin{itemize}
\item
Large enough internal volume,
\item
weak coupling,
\item
large complex structure values,
\item
and positivity of all the kinetic terms at the points of interest.
\item
Small periodicity of the axionic fields (a version of the Weak Gravity Conjecture),
\item 
sub-Planckian energy densities,
\item
and discrete values of the fluxes and the superpotential.
\end{itemize}

This is quite a long list of demands. It is not completely clear, {\it a priori}, that one could satisfy all of them with the limited number of parameters present in our model. We will show that it is indeed possible to overcome these difficulties and find examples that respect all these conditions.

There are, however, some approximations that we have made in order to simplify the problem. In particular, we have modelled the K\"ahler moduli sector of the compactification manifold by a single complex field. We have also taken a simple model for the non-perturbative superpotential whose field dependence is restricted to this single K\"ahler moduli. One could in principle perform the same kind of calculations in a more realistic version of our model with two K\"ahler moduli and two complex structure fields, as was done in ref.~\cite{Louis:2012nb, BlancoPillado:2006he}, to investigate if the results in our paper continue to hold in that case. Finally, we have not included any uplifting contribution in our examples.

After the main part of this work was completed, a new version of the de Sitter Swampland Conjecture appeared in ref.~\cite{Ooguri:2018wrx}.\footnote{Similar bounds on field theory potentials have been suggested by refs.~\cite{Dvali-bound.1,Dvali-bound.2,Andriot:2018wzk,Garg-bound.1,Garg-bound.2}, some of which appeared prior to ref.~\cite{Ooguri:2018wrx}.} This is a much weaker version of this conjecture that allows for saddle points in de Sitter space, but imposes some restrictions on the curvature of the potential at those critical points. The arguments behind this new version of the conjecture are different in nature to the previous one and they are being actively investigated at the moment~\cite{Hebecker:2018vxz}. Given this situation, we feel that it is still important to give concrete examples that can firmly establish whether any of these conjectures are valid in its current form. Therefore this work provides evidence that the strong version of the conjecture, as stated in eq.~(\ref{conjecture1}), is ruled out in string theory. We have also checked the form of the de Sitter saddle points we obtained in this model against the restrictions of the new conjecture, as reported in the final part of the paper.

The rest of the paper is organized as follows. In section~\ref{sec:overview}, we provide general details about string compactification models in Type IIB. In section~\ref{sec:examples}, we construct several explicit examples of potentials with de Sitter saddle points in this context. In section~\ref{sec:validity}, we discuss the validity of our solutions with respect to several possible constraints. In section~\ref{sec:inflation}, we investigate the form of the potentials around the de Sitter critical points and study them in connection to the refined version of the de Sitter Swampland Conjecture~\cite{Ooguri:2018wrx}. We conclude in section~\ref{sec:discussion}. Finally, some technical details of the compactification model as well as important input data for the specific examples are given in the appendices.

\section{Overview of Type IIB compactification scenarios}\label{sec:overview}

The most-studied models of string compactifications have been carried out using Type IIB orientifolds of Calabi-Yau (CY) threefolds with fluxes. The low energy effective theory in this case involves a set of moduli which are normally organized into the following categories: the complex structure moduli ($z_i$), the dilaton ($\tau$), and the K\"ahler moduli ($T_a$). Depending on the CY, the number of fields in these groups varies, but it could range up to $\mathcal{O}(100)$ fields. 

The dynamics of these fields at low energies is described by a four-dimensional $N=1$ supergravity model whose tree-level K\"ahler function is found to be
\beq
    K_{tree}(z_i,\tau,T_a) = -2 \log ({\cal V}) - \log (- i (\tau - \bar \tau)) - \log \left(-i \int_{\cal M}{\Omega \wedge \bar \Omega}\right)\,,
\eeq
where ${\cal V}$ is the volume of the internal dimensions in strings units and $\Omega$ denotes the holomorphic three-form of the CY manifold ($\mathcal{M}$).

Introducing fluxes along the internal cycles, one induces a potential for the complex structure and the dilaton, which can be computed using~\cite{Gukov:1999ya}
\beq
    W _{flux}(z_i, \tau) = \frac{1}{(2\pi)^2 \alpha'} \int_{\cal M}{( F_3 - \tau H_3) \wedge \Omega}\,,
\eeq
where $F_3$ and $H_3$ are the three-form fluxes that wrap around the 3d internal cycles. Taking into account these expressions for the K\"ahler potential and the superpotential, one obtains the $N=1$ F-term potential in Planckian units
\beq
    V_F = e^K \left(\sum_{A,B} K^{A \bar B} D_A W D_{\bar B} \bar W - 3 |W|^2\right)\,,
\label{eqn:scalar_potential}
\eeq 
where $A,B= \left(z_i,\tau,T_a\right)$ and we have denoted $D_A W = \partial_A W + W \partial_A K$.

Given the specific form of the K\"ahler moduli, one can show that the potential in this case becomes of the no-scale type, namely
\beq
V_F = e^K \left(\sum_{I,J} K^{I \bar J} D_I W_{flux} ~D_{\bar J} \bar W_{flux} \right)
\eeq
where $I, J = z_i, \tau$. It is clear from the form of the potential that we can find the minima for the complex structure and the dilaton by imposing the supersymmetric conditions
\beq
D_{z_i} W_{flux} =0\,, \qquad D_\tau W_{flux} =0\,.
\label{treelevelsusy}
\eeq
Evaluating the flux superpontential at this point will generically give a value different from zero, which we will denote as $W_0$. This would lead to supersymmetry breaking due to fluxes along the K\"ahler direction, since $D_{T_a} W_{flux} = W_0\partial_{T_a} K \neq 0$.

Finally, these vacua still have flat directions along the K\"ahler fields, so one must go beyond the no-scale limit in order to stabilize the K\"ahler moduli. This can be done either by introducing perturbative corrections to the K\"ahler function, or by adding non-perturbative terms to the superpotential. We will concentrate on non-perturbative terms in the simplest models of a single K\"ahler field, which take the form~\cite{Kachru:2003aw}
\beq
W_{np} = \sum_i A_i e^{-a_i T}\,.
\eeq
In the following, we will take $A_i$ and $a_i$ to be constants.

\section{Explicit examples }\label{sec:examples}

Let us now investigate the appearance of de Sitter critical points in a particular model with the ingredients described above. We will study the $N=1$ supergravity theory with the K\"ahler function
\beq
K(z_1,z_2,\tau,T) = -3 \log (T + \bar T) - \log (- i (\tau - \bar \tau)) + K_{cs}(z_1,z_2,\tau)
\eeq
and superpotential
\beq
W(z_1,z_2,\tau,T) = W_{flux}(z_1,z_2,\tau) + A e^{-a T} + B e^{-b T}\,.
\eeq
To improve the clarity of the presentation, we relegate to appendix~\ref{app:complex-structure} the specific functions for $K_{cs}(z_1,z_2,\tau)$, as well as $W_{flux}(z_1,z_2,\tau)$ and its dependence on the flux integers. Here, we have taken a model with two complex structure moduli, which is rich enough for our purposes. In fact, these functions are the appropriate ones for the well-studied orientifold model $\mathbb{P}^4_{11169}$~\cite{W0.1,manymin.1}. 

We use the ``racetrack-type superpotential'' in the non-perturbative correction. This has been argued to arise from gaugino condensation in a stack of $D7$ branes wrapped around some internal cycles of the CY geometry~\cite{Kachru:2003aw}. We take the form of the constant term in the exponent to be $2\pi/N$, with $N$ the rank of the associated gauge group.

This concludes the description of our model, which is characterized by several parameters that we will fix in the following examples. It is important to note that even when one fixes the field space manifold and the D-brane content of our compactification scenario, we will still have a large number of possible potentials available due to the multitude of possible fluxes. We will use this fact to show that our conclusions are quite generic.

\subsection{Supersymmetric vacua}

We start our description of potentials with de Sitter saddle points by studying an example of fluxes that give rise to a vanishing tree level flux superpotential, $W_0 = 0$.  Using only the complex structure moduli (meaning without introducing any non-perturbative terms), it has been shown in refs.~\cite{W0.1,W0.2} that such vacua are possible if one chooses the flux numbers, $(f_A|f_B)$ and $(h_A|h_B)$ (as defined in eq.~(\ref{eqn:fAB-hAB})), threading each cycle appropriately. For example, one can choose
\begin{equation}\label{eqn:fluxes-denef}
(f_A|f_B) = (20,0,0|0,-69,-28)\,,\qquad (h_A|h_B) = (0,-4,0|49,18,6)\,.
\end{equation}
to get $W_0=0$ at the solution of the supersymmetric eqs.~(\ref{treelevelsusy}). 

Adding the non-pertubative potential with parameters
\beq\label{eqn:racetrack-0}
A = -\frac{1}{100}\,, \qquad B = 1\,, \qquad a = \frac{2\pi}{100}\,, \qquad b = \frac{2\pi}{50}\,,
\eeq 
one can find a supersymmetric minima for all fields. In particular, we obtain $\text{Re}[T] = T_R = 82.430$ and $\text{Im}[T] = T_I =0$ at a AdS supersymmetric minima. It is important to note that we have solved the complete set of supersymmetric equations for all fields, so in fact the superpotential at the true minimum has a tiny component due to the small correction to the supersymmetric equations introduced by the non-perturbative terms. However, this correction of the position of the minima in field space in the complex structure and the dilaton is quite small. This is useful since it allow us to first solve the equations for the dilaton and the complex structure with $W_{np}=0$, and then use this solution as our initial guess for the full solution.

Note that the high rank of the gauge groups used here ($100$ for $a$ and $50$ for $b$) could lead to issues with backreaction on the internal manifold, due to many branes being stacked in the same place. This may lead to issues with our setup being realized in practice. See for example the discussion in ref.~\cite{Louis:2012nb}.

\subsection{de Sitter critical point}

\begin{figure}
\centering
\includegraphics[scale=0.75]{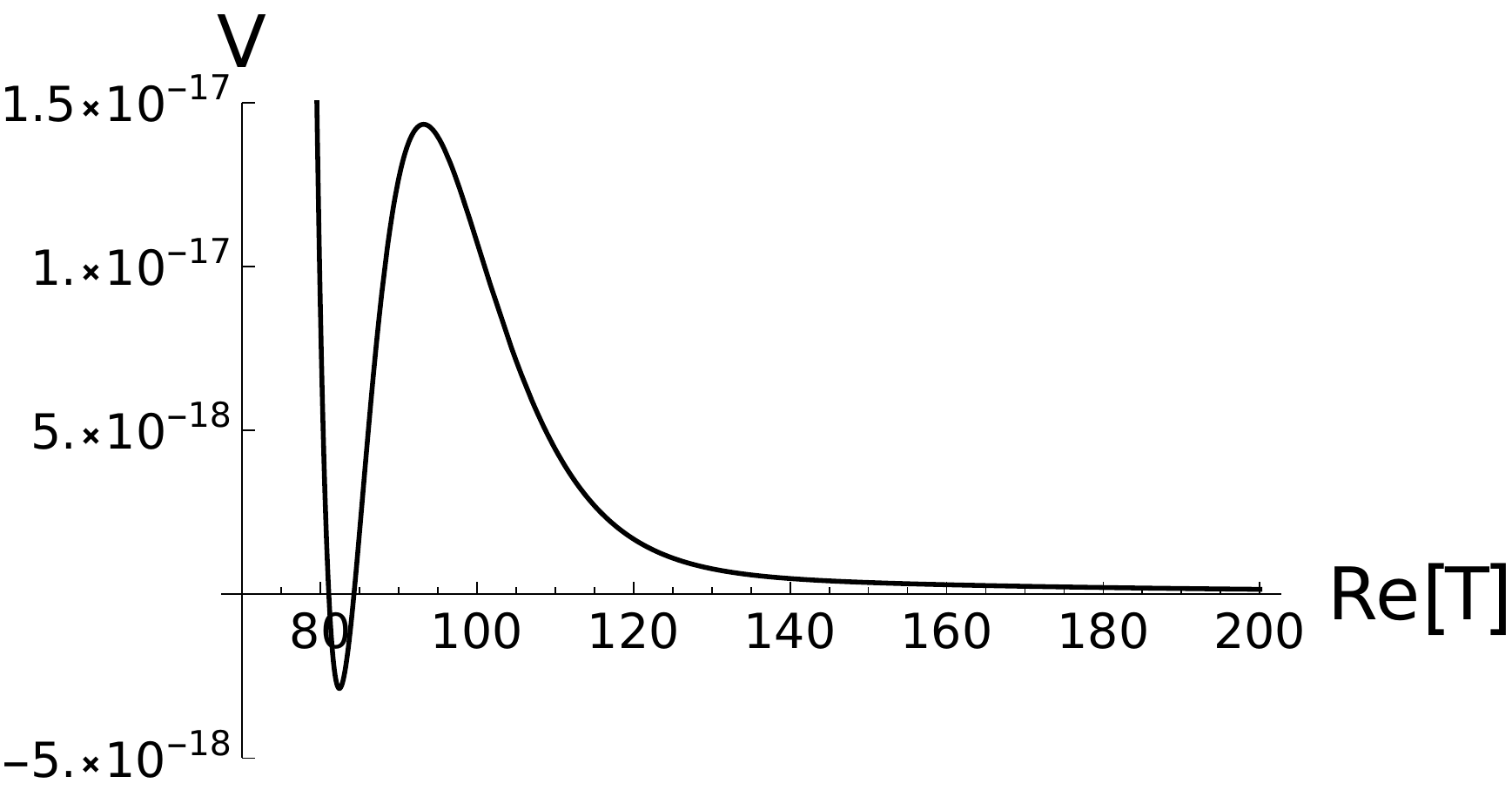}
\caption{A plot of the $W_0=0$ case (Case 0) using the racetrack parameters from eq.~(\ref{eqn:racetrack-0}). The nonperturbative correction results in an anti-de Sitter minimum. Because the potential asymptotically approaches zero from above, there must then be a maximum in $T_R$ between this minimum and infinity. That maximum is in fact very near to a de Sitter saddle point.}
\label{W0-zero}
\end{figure}

Looking at the asymptotic form of this potential at large values of $T_R$, one realizes that it approaches zero from above. However, the supersymmetric minimum we found before is at a negative value of $V$. This indicates, as figure \ref{W0-zero} shows, that the potential should have a local maximum at some intermediate value of $T_R$. This is the same idea described in ref.~\cite{Conlon:2018eyr} for the dilaton potential in a heterotic string compactification.

We have an expression for the scalar potential as a function of all the fields involved, so we can check that this maximum is indeed a critical point once we take into account all other directions. The locations of the dilaton and the complex structure moduli at the de Sitter critical point are slighly shifted from their values at the AdS supersymmetric vacuum. One can justify this by considering the differences in scales between the complex structure and dilaton masses and the K\"ahler fields. 

As noted above, the values of the fluxes in this example are such that the solution respects supersymmetry even before introducing any non-perturbative corrections. In other words, $W_0 = 0$. This makes this solution free of any of the potential problems described in ref.~\cite{Sethi:2017phn}, where some concern was raised about the introduction of non-perturbative terms in the superpotential without taking proper account the possibility of other perturbative corrections due to supersymmetry breaking. See, however, the discussion in ref.~\cite{Kachru:2018aqn}. 

We therefore conclude that it is possible to find true de Sitter saddle points in this type of scenario with many moduli fields. However, this example is somewhat special, since the main argument for the existence of the saddle point in the $T_R$ direction relies on the vanishing of the tree level flux superpotential. In the following, we relax this condition to see how generic de Sitter saddle points are in our models.

\subsection{More general cases}

In the previous section, we gave a particular example of the parameters that lead to the existence of a de Sitter critical point following the description given by Conlon~\cite{Conlon:2018eyr}.~\footnote{In ref.~\cite{Conlon:2018eyr}, Conlon argued how one can find other examples where these de Sitter saddle points can appear by using $\alpha'$ corrections to the potential. This requires a specific sign of these corrections. Here we will show that these points appear generically in the racetrack models even ignoring these corrections.} We will now show that such critical points exist for a large volume of the parameter space of the models we are using.

Let us start by describing another way in which one could try to find a de Sitter critical point in our construction. Consider the situation incorporating the non-perturbative terms such that they yield a supersymmetric Mikowski vacuum. In other words, we will consider the case where the total superpotential (not only $W_0$) at the vacuum is zero. This sounds like a good starting point if one wants to find a de Sitter saddle point: the potential around that minimum would be positive, but at large volume it should go back to zero, so it must turn around at some point. This, of course, only suggests the possibility of the existence of these points, and one will have to find explicit examples in the multifield potential. Here, we again use the $\mathbb{P}^4_{11169}$ CY complex structure moduli to give such examples. This model has only two complex structure moduli, but it is already rich enough to demonstrate the generic existence of de Sitter critical points in the Landscape.

Finding a supersymmetric Minkowski vacuum can be achieved with our racetrack potential, as was shown in refs.~\cite{Kallosh:2004yh,Blanco-Pillado:2005fn}, by adjusting the coefficients of our non-perturbative superpotential for a given $W_0$. In our case, we select\footnote{We again disregard issues of backreaction which would arise from these rank-300 and -150 groups condensing on the same cycle. Note that choosing low-rank groups still generically leads to de Sitter critical points, but at $T_R$ low enough to raise concerns about the supergravity approximation.}
\begin{equation}\label{eqn:racetrack-params}
    A = 0.26050-i0.30090\,,\qquad B=-0.65453+i0.75603\,,\qquad a=\frac{2\pi}{300}\,,\qquad b=\frac{2\pi}{150}\,,
\end{equation}
which were chosen based on a choice of fluxes
\begin{equation}\label{eqn:fluxes-minkowski}
    (f_A|f_B) = (20,-1,-6|12,-44,-12)\,,\qquad(h_A|h_B) = (-1,-4,3|43,21,7)\,.
\end{equation}
For these choices, we find an initial $W_0=-0.025920+i0.022994$. Looking at the full potential for these parameter values, we indeed find a de Sitter saddle point along the volume direction, as expected.

It may seem that this scenario is fine-tuned by the specific choice of our superpotential parameters $(A,B,a,b)$, such that we obtain the Minkowski vacuum. Thus, we shall vary the flux numbers while keeping the racetrack potential fixed, with the requirement that we only consider relatively small values of $|W_0|$ in our examples. We have scanned a few sets of flux integer values to identify a few suitable candidates for our purposes. We shall consider four cases: the one with $W_0=0$ from before, the one with fluxes as in eq.~(\ref{eqn:fluxes-minkowski}), and two others. These cases are detailed in Table~\ref{tbl:cases}.

\begin{table}
\centering
\begin{tabular}{|cccc|}
\hline
Case \# & $(f_A|f_B)$ & $(h_A|h_B)$ & $W_0$\\
\hline
0 & $(20,0,0|0,-69,-28)$ & $(0,-4,0|49,18,6)$ & 0 \\
1 & $(20,-1,-6|12,-44,-14)$ & $(-1,-4,3|43,21,7)$ & $-0.025920 + i0.022994$\\
2 & $(18,-2,-3|16,-37,-10)$ & $(-1,-4,3|46,21,5)$ & $-0.025987 + i0.000443$\\
3 & ~$(18,-1,-3|14,-42,-15)$~ & ~$(-1,-4,3|43,19,6)$~ & ~$-0.020426 + i0.011213$~\\
\hline
\end{tabular}
\caption{The flux integer choices and initial $W_0$ (found from solving $D_IW_{\text{flux}}=0$ for $I\in\{\tau,z_i\}$) for all four cases we studied. The values of eq.~(\ref{eqn:racetrack-params}) were chosen such that the full potential $V(z_i,\tau,T)$ of Case 1 has a Minkowski supersymmetric minimum.}\label{tbl:cases}
\end{table}

\begin{figure}
\centering
\includegraphics[scale=1]{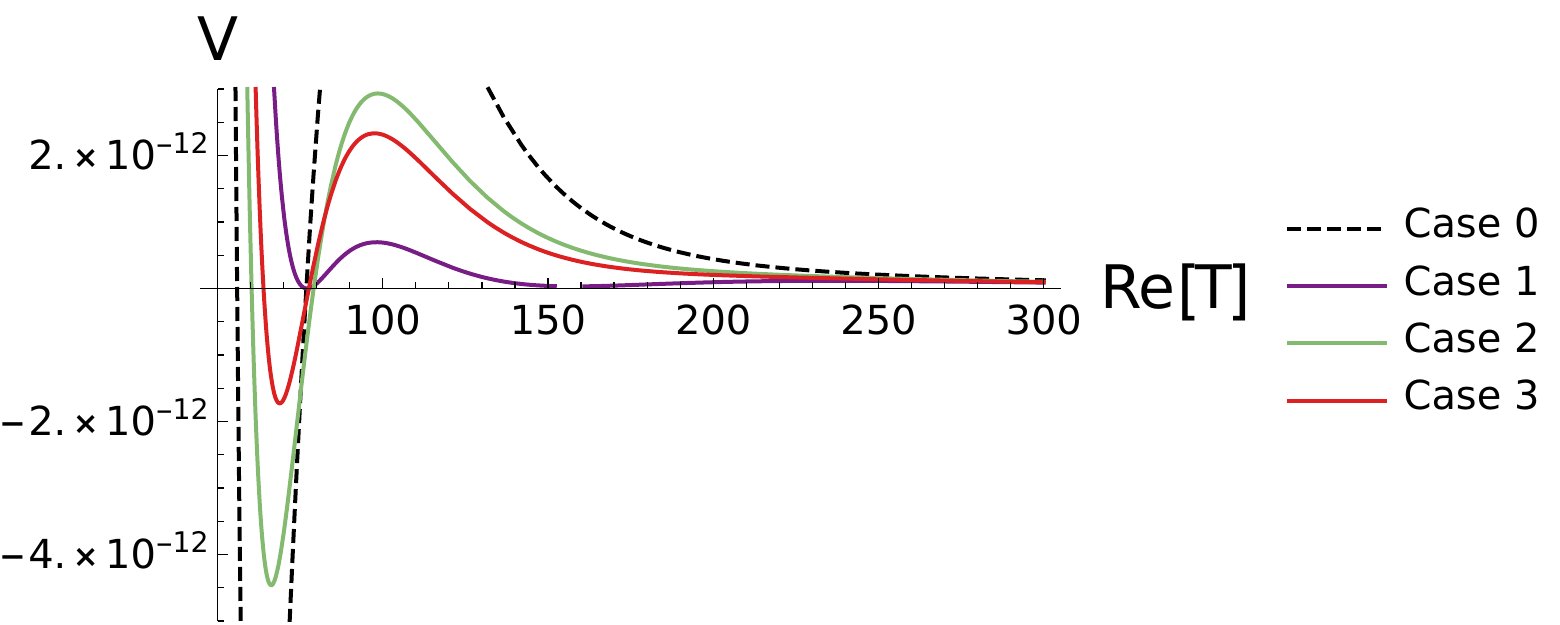}
\caption{A plot of the potentials of Cases 0--3 with $T_R$ varying and all other fields fixed at their first critical point values (vid. appendix~\ref{app:data}). All of the minima in this potential are anti-de Sitter, with the exception of Case 1, which is a Minkowski minimum. Cases 0, 2, and 3 are true minima.}
\label{SUSY1}
\end{figure}

We show in figure~\ref{SUSY1} the potential along the volume direction for all four cases, using always the racetrack parameters of eq.~(\ref{eqn:racetrack-params}). We find again a de Sitter critical point close to the minimum, where the potential is at an extremum in all field directions. There is another interesting point in this example for Cases 1--3. At large values of the volume, one finds another supersymmetric AdS critical point, so our de Sitter critical point is located in between these two supersymmetric points. This is a different asymptotic behavior than the one obtained Conlon~\cite{Conlon:2018eyr} and in our previous section. This fact makes it harder to see how can one modify the potential to avoid the de Sitter critical point without also destroying the nearby supersymmetric points. As before, there is typically a small shift in the values of the moduli between the supersymmetric minima and the de Sitter critical point.

Furthermore, Cases 2 and 3 exhibit the same general behavior as Case 1, and so such behavior does not seem to be the result of fine-tuning. In a realistic model with many complex structure moduli, the distribution of vacua in the $W_0$ around the origin is flat~\cite{Douglas:2004qg}, and the model we have chosen already allows for many minima around $|W_0|\approx0$~\cite{manymin.1,manymin.2}. This would mean that in reality, there would a very large number of these vacua with de Sitter critical points. Hence, the examples shown here are very generic in a typical CY.

\begin{figure}
\centering
\includegraphics[scale=1]{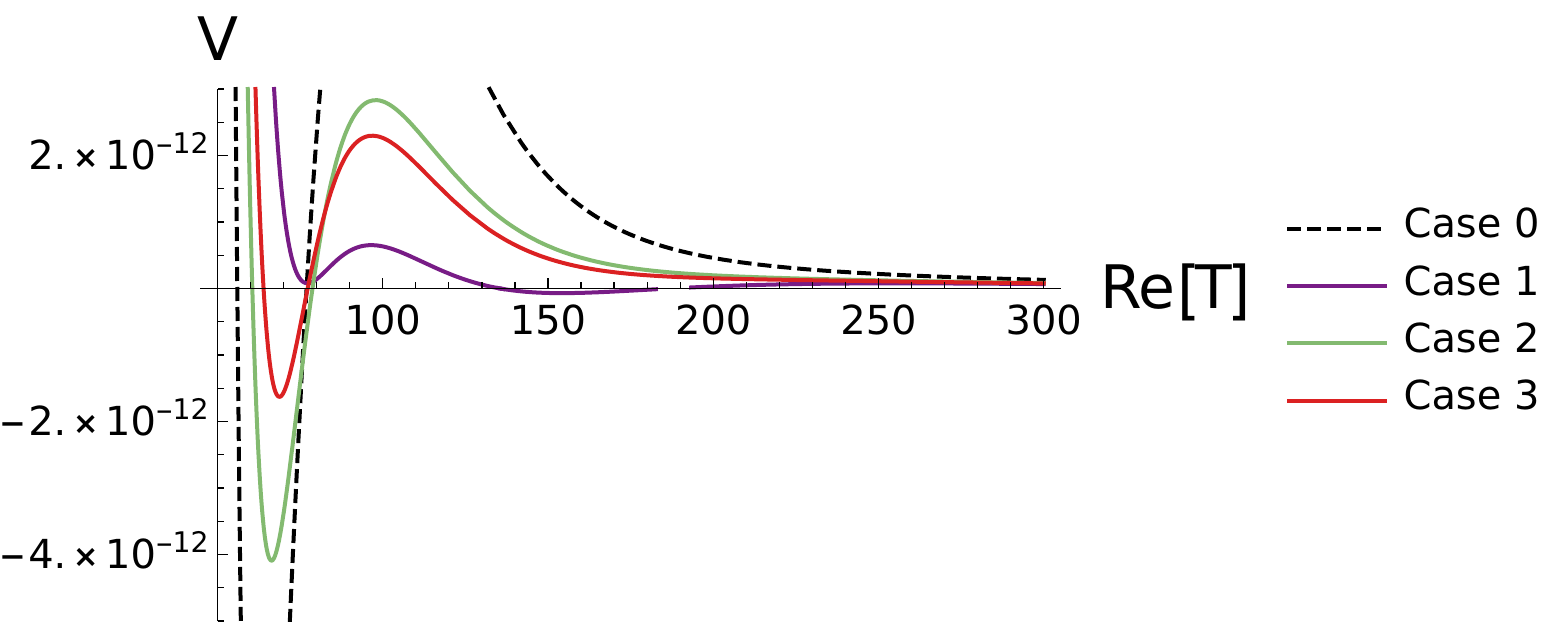}
\caption{A plot of the potentials of Cases 0--3 with $T_R$ varying and all other fields fixed at their de Sitter critical point values (vid. appendix~\ref{app:data}). All of the apparent maxima shown here are in fact saddle points once we account for all fields.}
\label{nSUSY}
\end{figure}

\begin{figure}
\centering
\includegraphics[scale=0.75]{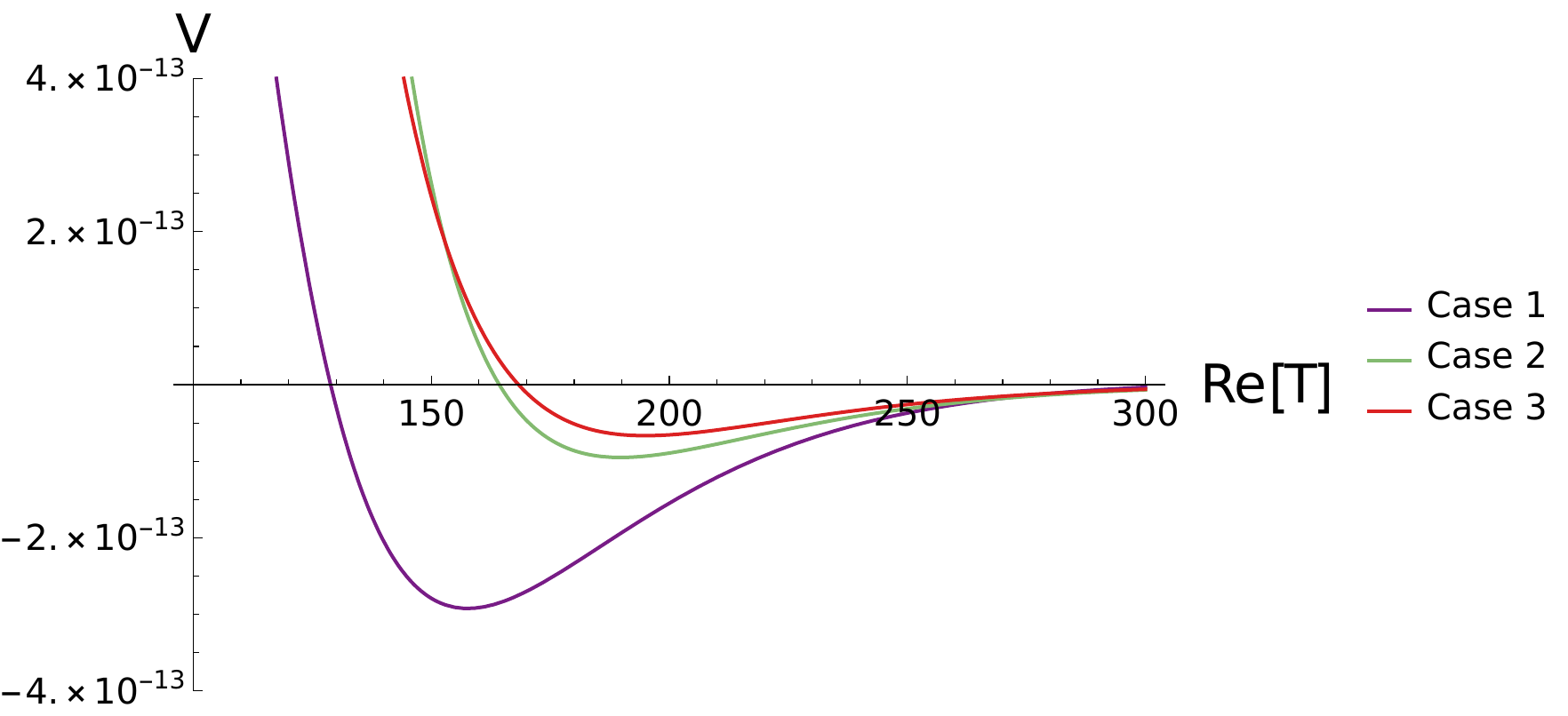}
\caption{A plot of the potentials of Cases 1--3 with $T_R$ varying and all other fields fixed at their second anti-de Sitter critical point values (vid. appendix~\ref{app:data}). All cases are true minima once we account for all the fields.}
\label{SUSY2}
\end{figure}

We plot in Figs.~\ref{SUSY1}~\&~\ref{nSUSY} the potentials along the volume direction for the values of the moduli that correspond to the first supersymmetric critical point and the de Sitter critical point. Finally, we plot in figure~\ref{SUSY2} the other supersymmetric AdS critical points that exist in all cases except the case with $W_0=0$. We give details about locations of these points in appendix~\ref{app:data}.

\section{Validity of the solutions}\label{sec:validity}

The values of the moduli at the minimum are constrained by several requirements so that one can trust the results given our approximations. The realizations we have studied here pass all of the following constraints.

One needs to find a minimum at a large value of the real part of $T$ so that the internal volume is large in string units and one can trust the supergravity approximation. Our smallest $T_R$ is $\approx62$.

The imaginary part of the dilaton should be large enough to be in the weak coupling regime. Similarly, the complex structure should be found in the region where the calculation of the periods that enter the K\"ahler and the superpotential can be trusted. Our smallest $\text{Im}[\tau]$ is $\approx 3$, and the $z_1$ all have imaginary part greater than $1$.

We also have to impose the positivity of all the kinetic terms for the moduli fields. This restricts some of the vacua that we have found numerically, forcing us to exclude them from consideration.

The values of the superpotential that we use have been obtained from a particular set of integer fluxes. In other words, we did not tune the value of the superpotential to fit our requirements. This imposes a serious constraint on the model since many of the possible vacua that one would find lie outside of the range of validity of our calculations.

Another important point that has been discussed in the literature recently~\cite{Moritz:2018sui} is the possible trans-Planckian periodicity of the axion fields associated with the K\"ahler moduli. Models with that property violate a generalization of the Weak Gravity Conjecture, and are therefore assumed to be part of the Swampland. This means that one should not consider such cases when looking for viable counterexamples of the de Sitter Swampland conjecture. 

This added restriction, together with all the other conditions we would like to satisfy, puts some tension on the set of possible parameters that one can use. However it is not hard to find examples where the periodicities for the axions are sub-Planckian. In fact, all the numerical examples we give in this paper avoid any violation of the Weak Gravity Conjecture. That is, we have
\begin{equation}
    aT_R>1\,,\qquad bT_R>1\,,\qquad (b-a)T_R>1
\end{equation}
for the $T_R$ at all critical points so all the terms in the potential for the axions have a sub-Planckian periodicity.~\footnote{Note that the dependence of this condition on $T_R$ comes from imposing the periodicity on the canonically normalized axion fields.} Finally, the de Sitter saddle points that we found are all at a sub-Planckian energy density, so this does not impose a serious restriction.

\section{de Sitter saddle points and the refined de Sitter conjecture}\label{sec:inflation}

In the previous sections, we have shown that it is quite generic to obtain de Sitter critical points in constructions of moduli stabilization without much fine tuning. These points rule out the strong version of the de Sitter Conjecture. 

A new version of the conjecture has appeared recently that allows de Sitter saddle points as long as the the curvature of the potential is large along the unstable direction~\cite{Ooguri:2018wrx}. It states that at a saddle point, the potential will satisfy the relation
\beq\label{conjecture2}
    \text{Min}\left( V_{\phi\phi}\right) \le - \frac{c'}{M_P^2} V\,,
\eeq
where $c'$ is a positive dimensionless constant of order $1$ and $V_{\phi\phi}$ denotes the second derivative of the potential with respect to the canonically normalized fields, $\phi$. This is directly related to one of the slow roll parameters in inflationary scenarios, $\eta$. This parameter is given by the ratio of the second derivative of the potential along the canonically normalized field direction and the potential itself. The new conjecture imposes that $\eta$ should be large and negative along the unstable direction at those de Sitter critical points.

Using the form of the potential for all the moduli fields we obtained earlier, one can find in our model the eigenvalues of the squared masses of the canonically normalized fields around any critical point~\footnote{See appendix~\ref{app:complex-structure} for the details of this calculation.}, and from there the values of $\eta$ at those points. This calculation shows that the unstable directions in all de Sitter saddle points are in fact pretty much aligned with the direction that corresponds to the volume modulus. This is in agreement with the na\"ive expectation one gets by looking at the plots of the potential along the $T_R$ direction.

We show in Table~\ref{tbl:etas} the values of $\eta$ for the four de Sitter critical points found earlier. It is clear from those results that all these points are in agreement with the weaker version of the de Sitter conjecture. However, this is to be expected in this simple model. The form of the kinetic term for this field is universal:
\beq
\frac{3}{4T_R^2}\left( \partial_{\mu} T_R \partial^{\mu} T_R\right)\,,
\eeq
and the value of the $\eta$ parameter for the canonically normalized field along the $T_R$ direction is in this case given by
\beq
\eta_{T_R} = \frac23 ~T_R^2 \left(\frac{V''}{V}\right)\,,
\eeq
where $V'' = \partial^2 V/\partial T_R^2$. Because we need to have a large volume for consistency of the model, we conclude that it would be very difficult to have small enough values of $\eta$. This is not surprising, since our model, at this level, is nothing more than a supergravity model and as such is likely to be affected by the so-called \emph{eta problem}~\cite{Copeland:1994vg}. In fact one can show that, without introducing new ingredients, the kind of model we have been discussing would not lead to a region of small $\eta$ parameter~\cite{Badziak:2008yg,Covi:2008cn}, no matter what numerical parameters we use for the model. However, it is clear that there are a number of possible extensions of this model that would allow for flat enough de Sitter critical points, possibly involving some fine tuning of the potentials along the axionic directions.

One example of this is given by the racetrack inflation models~\cite{BlancoPillado:2004ns,BlancoPillado:2006he} where an uplifting term was included in the discussion. It is quite remarkable that this simple modification allows for a realistic model of inflation to be implemented. We have not included such terms in this paper, as our focus was only to show that de Sitter critical points are generic in Type IIB compactifications, and thus we aimed to use a minimum number of ingredients.

\begin{table}
\centering
\begin{tabular}{|c|cccc|}
\hline
Case \# & 0 & 1 & 2 & 3 \\
~$\eta$ at de Sitter~ & $ \ -18.62 \  $ & $-38.58 \ $ & $-22.04 \ $ & $-22.20 \ $ \\
\hline
\end{tabular}
\caption{Values of the $\eta$-parameter for all cases at the de Sitter saddle points.}
\label{tbl:etas}
\end{table}

\section{Discussion}\label{sec:discussion}

In this paper, we have given explicit examples of de Sitter critical points in models of compactification with racetrack potentials. These points violate the de Sitter Swampland Conjeture given by eq.~(\ref{conjecture1}). We argue that in a generic CY with many moduli fields, there would be large numbers of these de Sitter critical points for generic values of the parameters of the racetrack superpotential and varying sets of flux numbers. We have shown this explicitly for a limited case, with only two complex structure moduli, to illustrate our point. For simplicity we have used a single-K\"ahler model, but we expect that one would be able to do the same exercise in the case of two K\"ahler~\cite{BlancoPillado:2006he}.

The de Sitter critical points found in our model satisfy the weaker version of the de Sitter Conjecture given by eq.~(\ref{conjecture2}). This is to be expected, given the nature of the unstable field direction and the fact that we only use a purely supergravity Lagrangian. However, given that the de Sitter Swampland Conjecture seems to be easily violated by these points, it would not be hard to envision cases where there will be flat enough saddle points once one introduces more ingredients to the Lagrangian, similar to what happens in models such as racetrack inflation~\cite{BlancoPillado:2004ns,BlancoPillado:2006he} .

We have also shown that one can find models of flux compactifications with many moduli which are fully compatible with all the constraints that one would normally like to impose in order to have control of the theory. In particular, we have shown that it is possible to find viable models of compactification that satisfy a version of the Weak Gravity Conjecture. This suggests that these types of models may be among the most interesting ones to find a de Sitter vacua in Type IIB compactifications. This has been studied in several scenarios in ref.~\cite{Kallosh:2014oja}, where de Sitter vacua are found in models with racetrack potentials of the kind discussed in refs.~\cite{Kallosh:2004yh,Blanco-Pillado:2005fn} and several uplifting mechanisms. 

\section{Acknowledgments}

We thank Igor Bandos, Jose L. F. Barbon, Mark Hertzberg, Renata Kallosh, Andrei Linde, Ander Retolaza, Kepa Sousa, Ra\"ul Vera, and Alex Vilenkin for useful conversations. We would like to further thank Kepa Sousa for collaboration on the early stages of this work. This work was supported in part by the Spanish Ministry MINECO grant (FPA2015-64041-C2-1P), and the Basque Government grant (IT-979-16). J. J. B.-P. is also supported in part by the Basque Foundation for Science (IKERBASQUE) and M.A.U. by the University of the Basque Country (PIF17/74).

\appendix

\section{Complex structure}\label{app:complex-structure}

As discussed in the text, we used the $\mathbb{P}^4_{11169}$ CY orientifold as a model for the complex structure moduli space. This model has been thoroughly discussed in the literature, viz~\cite{W0.1,manymin.1}, so we will restrict ourselves to a short discussion on the most important expressions we used to obtain the results described in the main text and appendix B. 

The prepotential arising from the model at hand, once the complex structure fields have been normalized, reads
\beq
	F=\frac{1}{6}\left( (9z_1^3 + 9 z_1^2 z_2^2 +3 z_1 z_2^2) - \frac{9}{4} z_1^2 - \frac{3}{2} z_1 z_2 - \frac{17}{4} z_1 - \frac{3}{2} z_2 + \xi \right) ~.
	\label{eqn:prepotential}
\eeq
Using the prepotential, we can compute the period vector $\Pi$:
\beq
\Pi^T = \left( \ 1 \ , \  z_1 \ , \ z_2 \ , \ 2F - z_1 F_1 - z_2 F_2 \ , \ F_1 \ , \ F_2 \ \right)
\label{eqn:period}
\eeq
where $F_i = \partial_{z_i} F$. The K\"ahler potential involving the complex structure moduli is thus obtained via
\beq 
K_{cs} (z_1, z_2) = - \log \left( i \Pi^\dag \cdot \Sigma \cdot \Pi \right)
\label{eqn:Kcs}
\eeq
where $\Sigma$ is the 6-by-6 symplectic matrix. In our case, plugging (\ref{eqn:period}) into (\ref{eqn:Kcs}), we get
\beq
K_{cs} (z_1, z_2) = - \log \left( 4 Y_1 (3 Y_1^2 + 3 Y_1 Y_2 + Y_2^2) - 4 i \xi \right)
\eeq
where $z_i = X_i + i Y_i$ and  $\xi = -1.3 i$ .
\\

Along with the K\"ahler potential, the key piece to obtain the scalar potential is the superpotential. Fluxes threading the internal space yield a contribution to the superpotential~\cite{Gukov:1999ya} given by
\beq
W_{flux} = \frac{1}{(2\pi)^2 \alpha'} \int_{\cal M} (F_3 - \tau H_3) \wedge \Omega
\eeq
where the integral is performed over the whole internal manifold $\cal M$, $F_3$ and $H_3$ are 3-form fluxes present in type IIB String Theory and $\Omega$ is the holomorphic 3-form of the CY.  

Choosing a symplectic basis of the three-cycles of $\cal M$, $\lbrace A_i , B_i \rbrace$ $(i=0,1,2)$, and taking into account the flux quantization conditions
\beq\label{eqn:fAB-hAB}
f_{A,B}^i = \frac{1}{(2\pi)^2 \alpha'} \int_{A^i,B^i} F_3 \in \mathbb{Z}, \qquad h_{A,B}^i = \frac{1}{(2\pi)^2 \alpha'} \int_{A^i,B^i} H_3 \in \mathbb{Z}
\eeq 
the flux superpotential can be shown to be 
\beq
W_{flux} = \sum_{i=0}^{2} \left[ (f^i_A - \tau h^i_A) F_i - (f_B^i - \tau h_B^i) z_i \right] = N^T \cdot \Sigma \cdot \Pi
\eeq
where $F_0 = 2F - z_1 F_1 - z_2 F_2$ and we have defined 
\beq
N^T = \left( \ f_0 - \tau h_0 \ , \ f_1 - \tau h_1 \ , \  f_2 - \tau h_2 \ \right).
\eeq

Once $K$ and $W_{flux}$ have been computed, the scalar potential $V(\tau,z_1,z_2,\rho)$ is obtained from (\ref{eqn:scalar_potential}). As mentioned in the text, we will be interested in obtaining the eigenvalues of the Hessian of $V$ so we can compute the $\eta$ parameter at that point. We cannot, however, compute the Hessian matrix as it stands. Looking at the $\mathcal{N}=1$ supergravity action
\beq
S = - \int d^4 x \sqrt{-g} \left[ \frac{1}{2} R + K_{I \bar{J}} \partial_\mu \Phi^I \partial^\mu \bar{\Phi}^{\bar{J}} + V(\Phi^M, \bar{\Phi}^{\bar{M}}) \right]
\eeq
where $\Phi^I= \lbrace \tau, z_1 , z_2, \rho \rbrace$, we note that the kinetic term is not in canonical form. Moreover, we will need to compute the eigenvalues with respect to the real and imaginary parts of all the moduli, so we will need to find a metric $g_{ij}$ in field space such that
\beq
K_{I \bar{J}} \partial_\mu \Phi^I \partial^\mu \bar{\Phi}^{\bar{J}} = \frac{1}{2} g_{ij} \partial_\mu \phi^i \partial^{\mu} \phi^j
\eeq
where $\phi^i$ stands for the real or imaginary part of any moduli. Thus, the matrix from which the eigenvalues will be computed is 
\beq
\mathcal{H}^i_{\phantom{j}j}=g^{ik} (\partial_k \partial_j V - \Gamma^l_{kj} \partial_l V).
\eeq
Note that the second term will vanish if we are analyzing a critical point.  In the following section we give some results for these eigenvalues at critical points. Using these eigenvalues one can compute the value of the slow roll parameter $\eta$ at the de Sitter saddle points. The results of this calculation are shown in Table~\ref{tbl:etas}.

\section{Data for examples}\label{app:data}

Here we present the values of $z_1$, $z_2$, $\tau$, and $T$ at the various critical points we illustrate in Figs.~\ref{SUSY1}-\ref{SUSY2}. While we report these values to five significant digits, all calculations were carried out to a precision of forty digits. While it is not always clear from these tables, owing to the low precision in the presentation, the $\tau$, $z_1$, and $z_2$ values all change between critical points, and so these data should be taken to give the vicinity of the critical points.

At the first (lower $T_R$) supersymmetric critical points, the fields take the following values:

\begin{center}
\begin{tabular}{|crrrr|}
\hline
Case \# & $\hfill\tau\hfill$ & $\hfill z_1 \hfill$ & $\hfill z_2 \hfill$ & $\hfill T \hfill$ \\
\hline
0 & $\,\,-8.712\text{E-4}+i3.001$ & $\,\,-1.108\text{E-4}+i1.000$ & $\,\,9.850\text{E-5}+i0.9999$ & $\,\,62.22-i1.044\text{E-3}$ \\
1 & $-0.8120+i3.752$ & $-0.5176+i1.383$ & $ 1.067+0.1546$ & $77.09+i0$ \\
2 & $-0.5595+i3.395$ & $-0.6514+i1.225$ & $ 1.543+i0.6160$ & $66.25+i5.737$ \\
3 & $-0.5748+i3.485$ & $-0.4232+i1.304$ & $ 1.011+i0.4240$ & $68.82+i2.727$ \\
\hline
\end{tabular}
\end{center}
At the non-supersymmetric de Sitter saddle points, the fields take the following values:

\begin{center}
\begin{tabular}{|crrrr|}
\hline
Case \# & $\hfill\tau\hfill$ & $\hfill z_1 \hfill$ & $\hfill z_2 \hfill$ & $\hfill T \hfill$ \\
\hline
0 & $\,\,-8.929\text{E-4}+i3.001$ & $\,\,-1.136\text{E-4}+i1.000$ & $\,\,1.010\text{E-4}+i0.9999$ & $\,\,96.53+i5.127\text{E-4}$\\
1 & $-0.8118+i3.752$ & $-0.5176+i1.383$ & $1.067+i0.1546$ & $96.53+i1.178\text{E-3}$\\
2 & $-0.5593+i3.395$ & $-0.6514+i1.225$ & $1.543+i0.6159$ & $97.84+i3.846$\\
3 & $-0.5747+i3.485$ & $-0.4232+i1.304$ & $1.011+i0.4240$ & $96.85+i1.963$\\
\hline
\end{tabular}
\end{center}

At the second (higher $T_R$) supersymmetric anti-de Sitter minima, the fields take the following values:

\begin{center}
\begin{tabular}{|crrrr|}
\hline
Case \# & $\hfill\tau\hfill$ & $\hfill z_1 \hfill$ & $\hfill z_2 \hfill$ & $\hfill T \hfill$ \\
\hline
1 & $\,\,-0.8110+i3.751$ & $\,\,-0.5175+i1.383$ & $\,\,1.067+i0.1546$ & $\,\,157.6+i3.882\text{E-3}$ \\
2 & $-0.5584+i3.395$ & $-0.6513+i1.225$ & $1.543+i0.6158$ & $189.7-i43.34$ \\
3 & $-0.5738+i3.484$ & $-0.4231+i1.304$ & $1.010+i0.4240$ & $194.9-i18.37$ \\
\hline
\end{tabular}
\end{center}

Finally, at the critical points, the mass spectra of the scalar fields is the following:

\begin{center}
\tiny
\begin{tabular}{|ccc@{\hskip10pt}c@{\hskip10pt}c@{\hskip10pt}c@{\hskip10pt}c@{\hskip10pt}c@{\hskip10pt}c@{\hskip10pt}c|}
\hline
Point  & ~Case \#~ & \multicolumn{8}{c|}{Masses} \\
\hline
\multirow{4}{*}{1st SUSY} & 0 & $5.791\text{E-4}$ & $5.790\text{E-4}$ & $2.846\text{E-4}$ & $2.845\text{E-4}$ & $1.510\text{E-4}$ & $1.510\text{E-4}$ & $9.744\text{E-10}$ & $8.462\text{E-10}$ \\
         & 1 & $1.015\text{E-2}$ & $1.015\text{E-2}$ & $3.174\text{E-4}$ & $3.174\text{E-4}$ & $1.033\text{E-4}$ & $1.033\text{E-4}$ & $9.125\text{E-11}$ & $9.125\text{E-11}$ \\
         & 2 & $1.813\text{E-3}$ & $1.813\text{E-3}$ & $4.155\text{E-4}$ & $4.154\text{E-4}$ & $1.523\text{E-4}$ & $1.523\text{E-4}$ & $4.292\text{E-10}$ & $3.799\text{E-10}$ \\
         & 3 & $1.836\text{E-3}$ & $1.836\text{E-3}$ & $3.991\text{E-4}$ & $3.991\text{E-4}$ & $1.316\text{E-4}$ & $1.316\text{E-4}$ & $2.889\text{E-10}$ & $2.636\text{E-10}$ \\
         \hline
\multirow{4}{*}{de Sitter saddle } & 0 & $1.551\text{E-4}$ & $1.550\text{E-4}$ & $7.622\text{E-5}$ & $7.614\text{E-5}$ & $4.045\text{E-5}$ & $4.040\text{E-5}$ &  \color{red}{\boldmath$-1.269\text{E-10}$} & $1.178\text{E-10}$\\
         & 1 & $5.169\text{E-3}$ & $5.168\text{E-3}$ & $1.616\text{E-4}$ & $1.616\text{E-4}$ & $5.262\text{E-5}$ & $5.260\text{E-5}$ & \color{red}{\boldmath$-2.514\text{E-11}$} & $2.427\text{E-11}$ \\
         & 2 & $5.630\text{E-4}$ & $5.628\text{E-4}$ & $1.290\text{E-4}$ & $1.289\text{E-4}$ & $4.730\text{E-5}$ & $4.726\text{E-5}$ & \color{red}{ \boldmath$-6.244\text{E-11}$} & $5.866\text{E-11}$ \\
         & 3 & $6.590\text{E-4}$ & $6.589\text{E-4}$ & $1.432\text{E-4}$ & $1.432\text{E-4}$ & $4.723\text{E-5}$ & $4.719\text{E-5}$ &  \color{red}{\boldmath$-5.101\text{E-11}$} & $4.795\text{E-11}$ \\
         \hline
\multirow{3}{*}{2nd SUSY} & 1 & $1.188\text{E-3}$ & $1.188\text{E-3}$ & $3.713\text{E-5}$ & $3.713\text{E-5}$ & $1.209\text{E-5}$ & $1.208\text{E-5}$ & $1.294\text{E-12}$ & $9.632\text{E-13}$ \\
         & 2 & $7.720\text{E-5}$ & $7.720\text{E-5}$ & $1.769\text{E-5}$ & $1.769\text{E-5}$ & $6.484\text{E-6}$ & $6.483\text{E-6}$ & $3.801\text{E-13}$ & $3.065\text{E-13}$ \\
         & 3 & $8.087\text{E-5}$ & $8.087\text{E-5}$ & $1.757\text{E-5}$ & $1.757\text{E-5}$ & $5.793\text{E-6}$ & $5.792\text{E-6}$ & $2.697\text{E-13}$ & $2.172\text{E-13}$ \\
         \hline
\end{tabular}
\end{center}
The eigenvectors corresponding to the last two masses of every row are almost aligned with $T_R$ and $T_I$, respectively. Eigenvalues corresponding to the $T_R$ direction for the de Sitter critical point have been highlighted in red.

\end{document}